\documentclass[reprint,superscriptaddress, amsmath,amssymb, floatfix, prb]{revtex4-1}
\usepackage{xcolor}
\usepackage{graphicx}
\usepackage{braket}
\usepackage{mathtools}
\usepackage[ ]{hyperref}
\usepackage[utf8]{inputenc}
\usepackage[T1, OT1]{fontenc}
\usepackage{dsfont}
\usepackage{lipsum}
\usepackage[normalem]{ulem}

\begin{document}

\title{Non-Markovianity between site-pairs in FMO complex using \\
discrete-time quantum jump model}


\author{Mousumi Kundu}
\affiliation{ Indian Institute of Science Education and Research, Berhampur}
   
\author{C. M. Chandrashekar}
\affiliation{Quantum Optics \& Quantum Information,  Department of Instrumentation and Applied Physics, Indian Institute of Science, Bengaluru 560012, India}
\affiliation{The Institute of Mathematical Sciences, C. I. T. Campus, Taramani, Chennai 600113, India}
\affiliation{Homi Bhabha National Institute, Training School Complex, Anushakti Nagar, Mumbai 400094, India}

\begin{abstract}
\begin{center}
\bf{Abstract}
\end{center}
The Fenna-Mathews-Olson (FMO) complex present in green sulphur bacteria is known to mediate the transfer of excitation energy between light-harvesting chlorosomes and membrane-embedded bacterial reaction centres. Due to the high efficiency of such transport process, it is an extensively studied pigment-protein complex system with the eventual aim of modelling and engineering similar dynamics in other systems and use it for real-time application. Some studies have attributed the enhancement of transport efficiency to wave-like behaviour and non-Markovian quantum jumps resulting in long-lived and revival of quantum coherence, respectively.  Since dynamics in these systems reside in the quantum-classical regime, quantum simulation of such dynamics will help in exploring the subtle role of quantum features in enhancing the transport efficiency, which has remained unsettled.  Discrete simulation of the dynamics in the FMO complex can help in efficient engineering of the heat bath and controlling the environment with the system. In this work, using the discrete quantum jump model we show and quantify the presence of higher non-Markovian memory effects in specific site-pairs when internal structures and environmental effects are in favour of faster transport. As a consequence, our study leans towards the connection between non-Markovianity in quantum jumps with the enhancement of transport efficiency.
\end{abstract}

\maketitle
\section{Introduction}
 Photosynthesis is one of the primordial processes in nature that provides energy to sustain life.  The dynamics of photosynthetic organisms, which have successfully harvested and transferred solar energy for several billion years, is far more efficient than any man-made device known to date\,\cite{initial_1,initial_3, initial_4,initial_5}. Decades of studies have attributed the remarkable efficiency of the photosynthesis process to light-harvesting antenna complexes that are funnelling the excitation energy from captured photons to a reaction centre, where it is converted to chemical energy. Photosynthetic complexes have differences based on their living conditions and habitats, but all follow the same procedure: they absorb solar energy (photons) in the form of electronic excitation by an antenna, and then this excitation is transported to a reaction centre where charge separation transforms it into a more stable form of energy\,\cite{initial_6,initial_7}. The most simple and well-studied examples of such a light harvesting system are found in green sulphur bacteria, which is an organism that depends only on sunlight as a source of energy. Its antenna is quite large and made out of chlorosomes, which allows them to thrive in extremely low light conditions. A special kind of structured complex can be found in them, called the Fenna-Matthews-Olson (FMO) complex. These complexes connect the antenna to the reaction centre\,\cite{initial_8,initial_9}. They are small in size, water-soluble, and, most importantly, they transport excitation from the antenna to the reaction centre with more than $90\%$ efficiency. Even though these excitons are short-lived with lifetime of less than a nanosecond, the efficiency of the transport process has motivated and encouraged intense research in the direction of understanding and modelling environment assisted energy transport\,\cite{6, 7, 8, review_qc}. 
 
Mathematical modelling to mimic the observed dynamics is one of the most effective ways to understand the dynamics in these photosynthetic complexes.  In 2007, a new model was presented with evidence for wavelike behaviour and presence of quantum coherence during these exciton transport\,\cite{initial_11} . Photosynthesis usually occurs at ambient temperature, and the presence of quantum coherence at such temperature for $300 - 500$ fs led to numerous researches in the direction of exploring environment-assisted quantum transport to find further evidence for the presence of quantum coherence for high transport efficiency\,\cite{entanglement}.  Numerous studies have outlined the structural details of the FMO complex and that has been used to study and understand the dynamics in quantum-mechanical framework\,\cite{review_fmo}.  In the widely accepted form, the FMO complex is a trimer formed by three identical monomers that each bind seven Bacteriochlorophyll-a (BChla) molecules, since monomers function independently, without loss of generality, studies are restricted to a single monomer.  There are seven sites in the FMO complex, and it is assumed that there is at most one exciton in the complex at any time. Considering one exciton at any time is a reasonable assumption because these bacteria usually receive very less sunlight. The initial excitation occurs at site 1 or 6 and is transported to the sink at sites 3 and 4, the two dominant pathways for exciton transfer are $(1 \rightarrow 2 \rightarrow 3)$ and $(6 \rightarrow (5, 7) \rightarrow 4 \rightarrow 3)$ \cite{nature_fmo}. We would also like to note that the dynamics in FMO complex is also studied as a dimer using Bloch-Redfield equations\,\cite{dimer}. In this work we will stick to FMO complex as a trimer formed by identical monomers.

In the wavelike description of dynamics in the FMO complex which has put forward the contribution of quantum coherence for better efficiency, the wave function of the excitation enters a superposition state of multiple combined pigments instead of the excitation transferring from pigment to pigment sequentially.  The wave function enters superposition state  due to the strong electronic couplings between chromophores resulting in de-localized exciton state.  This process enable the rapid and coherent transfer of excitation along multiple paths at once and find the shortest route from the antenna to the reaction centre\,\cite{coherence}. A more rigorous and generic theoretical treatment of the FMO complex in quantum mechanical framework has been presented in two seminal works by Plenio \& Huelga\,\cite{initial_17} and Mohseni et al.\,\cite{base}, respectively. There, it has been shown that only coherence dynamics cannot transport exciton with such high efficiency, and the interplay between coherent dynamics and environmental induced noise transports exciton with such high efficiency. For these studies, the FMO complex and its surrounding environment have been modelled into an open quantum system set-up, and quantum master equations have been used to solve these problems\,\cite{opqm1,opqm2}. 

Studies have also shown that the coherence might be helpful in transport dynamics, but it does not contribute to the high efficiency of this exciton energy transport dynamics\,\cite{coherence1}. An other recent study argue that the contribution from quantum coherence is minute at best in photosynthesis transport process\,\cite{ZY21}. Other than quantum coherence, quantum features like entanglement and quantum memory effects (non-Markovianity) have been explored additionally to find their impacts on this highly efficient transport process. The presence of bipartite entanglement in the dynamics of FMO complex has been reported\,\cite{nature_fmo} and it has been tried to understand its usefulness in quantum technologies. Studies have also reported the role of non-Markovianity 
these types of efficient transport process\,\cite{non_mar, MMR20}. In spite of all these studies, the role of quantum features and quantum advantage in photosynthesis continues to remain unsettled paving way for further investigations.

In addition to all these studies, quantum simulation of the dynamics in FMO complex has also been explored. In quantum simulation approach, one can engineer the dynamics and control the parameters that could effectively lead to the observed phenomena and improve our understanding of the complex properties in FMO system.  With noisy intermediate-scale quantum (NISQ) devices\,\cite{new} being available, FMO complex systems which are defined on 7-site system can be simulated on a smaller number of qubit, NISQ devices. Quantum simulators are available in two different formats: analog quantum simulators and digital quantum simulators. Analog quantum simulators use continuous-time evolution equations and the evolution protocol has to be explicitly worked for each system on which it is simulated. Digital quantum simulations use universal quantum gates to simulate the discrete-time evolution equation, thus it provides flexibility and universality over analog quantum simulation\,\cite{dig_qm,analog}. Analog quantum simulations of the FMO complex have been reported using ultracold atoms\,\cite{ultra}, superconducting circuits\,\cite{27, 28} and NMR quantum computer\,\cite{29, 30}. A setup for digital simulation of the FMO complex has also been reported, but their study does not include the interplay of coherent dynamics and quantum jumps\,\cite{32}. Recently, a framework for digital quantum simulation for FMO complex has been presented which explicitly includes the interplay of coherent dynamics and quantum jumps\,\cite{base1} and a general algorithm form of FMO complex dynamics in open quantum dynamics framework has also been reported\,\cite{quantum_6}.  Using the controllable parameters in the discrete-time evolution framework composing of an interplay of coherent dynamics and quantum jumps\,\cite{base1}, one can quantify non-trivial quantum features in the dynamics and have a better understanding of the high-efficient exciton transport process of FMO complex.

In this work, using the discrete-time quantum jump model to study  dynamics in FMO complex  we study the dynamics between site pairs and quantify the non-trivial quantum features which may aid the high transport efficiency.  For modelling the quantum dynamics explicitly between site-pairs, we have prepared a theoretical framework by adopting the conceptual understanding from the previous generic framework where all sites are considered as the system and protein environment as the bath\,\cite{base1}.  For our study, only the the specific site pair is considered as system and remaining sites along with protein environment is treated as bath. In particular,  we have studied the non-Markovian memory effects in the dynamics between different site-pairs of FMO complex.  To identify and quantify the non-Markovian memory effects in the dynamics we have used trace distance and Breuer-Laine-Piilo (BLP) measure, respectively.  BLP measure is based on the quantification of the flow of information between the open system and its environment using rate of change of trace distance. In BLP measure, the information flow is the rate of change of the trace distance between the quantum states\,\cite{review_nm}.  The numerical simulation results shows the presence and quantifies the non-Markovian memory effects in some specific site pairs. This memory effect is controlled by the structural features of the FMO complex (site couplings and site energy differences) and environmental influence.  This result can be useful for studying finer details of photosynthetic complex. Our results match with previous theoretical findings of presence of non-Markovian memory effect and explicitly quantify the non-Markovian memory effects in those conditions where internal structures and environmental effects are in favour of faster transport.  Our approach using site-pair and bath model signifies that digital simulation method provides flexibility in controlling the dynamics and is universal and implementable in near-term device.

 This article is organised as follows, in Sec.\,\ref{model_methods} we present the model with open quantum system framework to model the system-environment coupling,  study site-pair dynamics and method to quantify the non-Markovian behavior in the dynamics.  In Sec.\,\ref{result} we presents the results for different pairs of sites in the FMO complex and analyze non-Markovian behaviour in the dynamics and in Sec.\,\ref{conc} we conclude with the summary of our observations. 

\section{Model and Methods}\label{model_methods}

The FMO complex serves as a transport channel in green sulphur bacteria which transports the exciton energy from site 1 or site 6 to site 3. The dynamics of FMO complex systems is championed by quantum features and can be modeled into open quantum system dynamics. Thus, its quantum and dissipative effects can be captured using a combination of Kraus operators and unitary quantum evolution. By using those discrete time evolution operators the exciton energy transport in 7-site FMO complex can be effectively and efficiently simulated using the digital quantum simulation framework\,\cite{base1}. Thus, in this work we use the previously proposed open quantum system framework of energy transfer in 7 site FMO complex to prepare a simulation framework for site-pairs (2-site system). 
\begin{figure}[!h]
\centering
\includegraphics[width=0.23\textwidth]{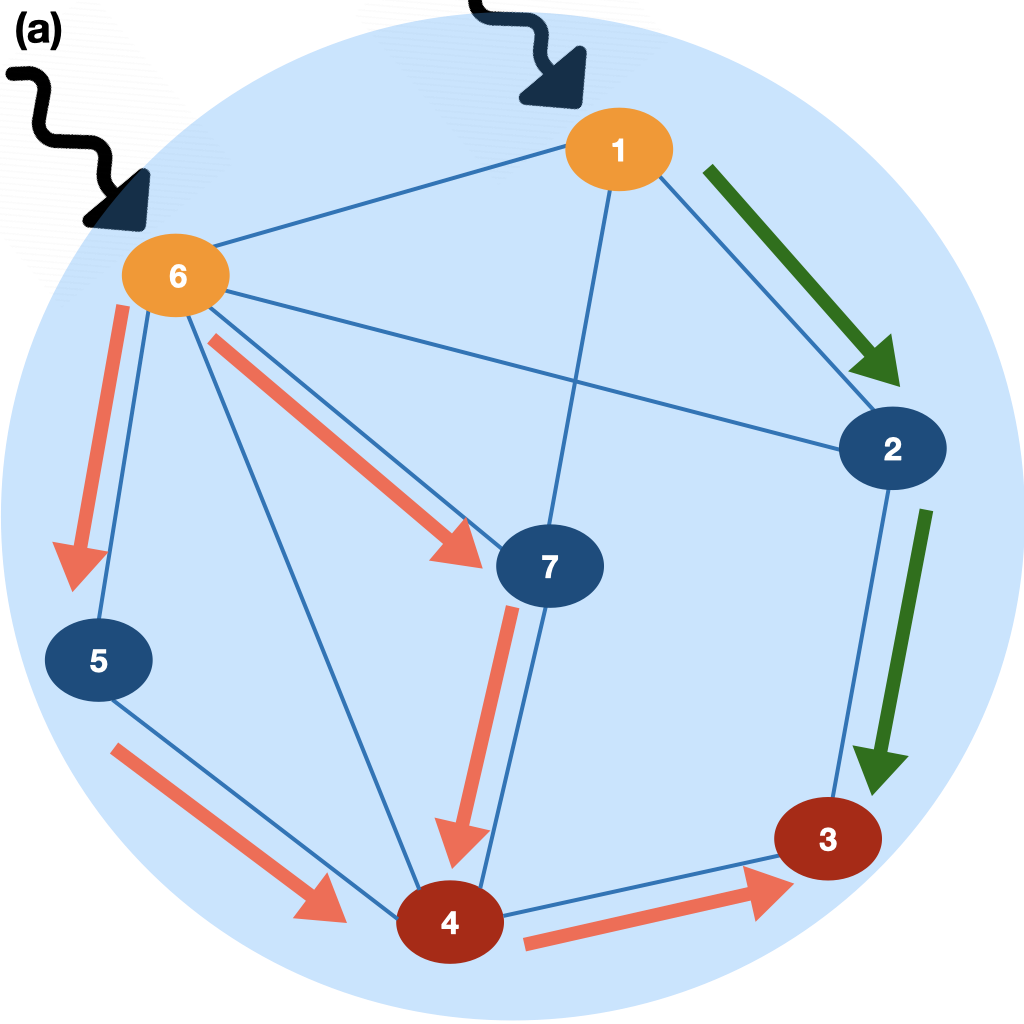}
\includegraphics[width=0.23\textwidth]{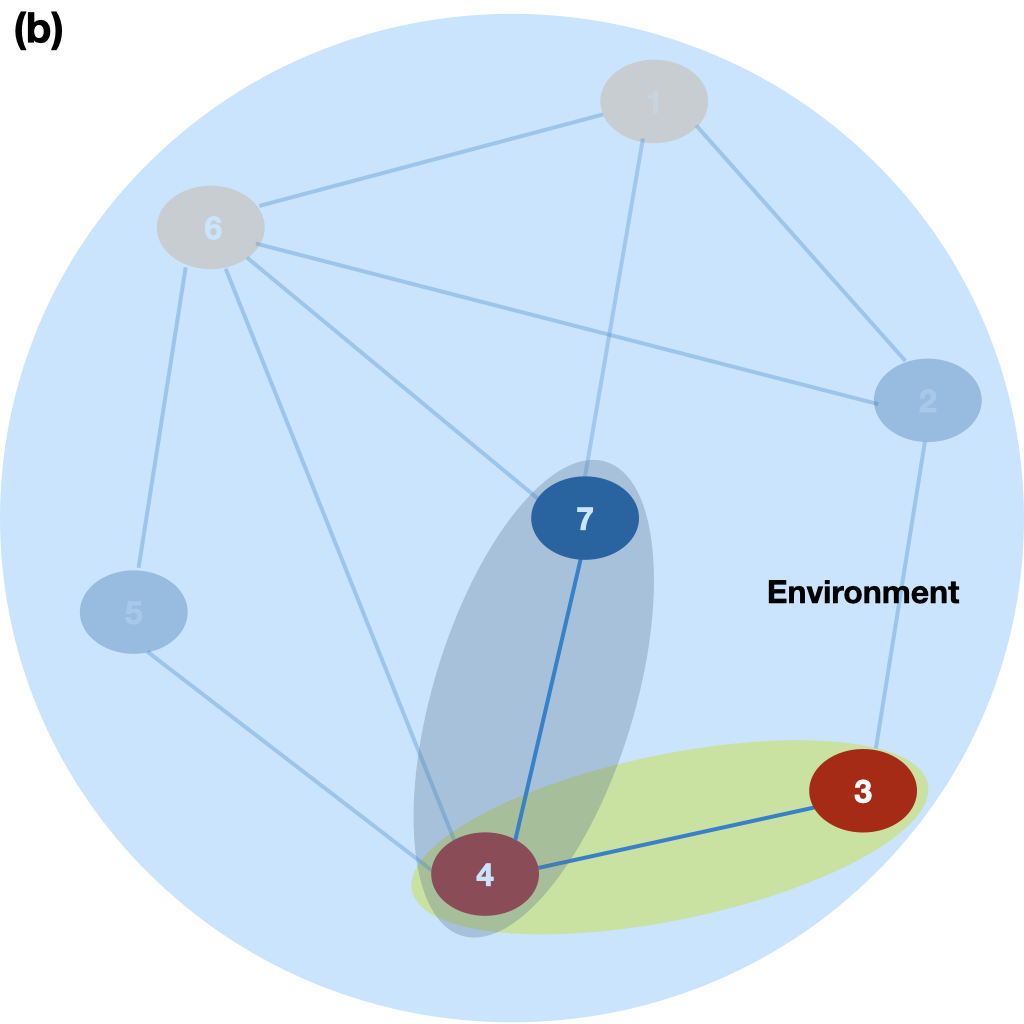}
\caption{\label{fig:figure1}(a) A schematic structure of 7-site FMO complex where each sphere represents a BChla site. Site 1 and 6 represent the initial site where excitation occurs. Site 3 and 4 represent the sink site to which excitation gets transported though intermediate sites, 2, 5, and 7. (b) Schematic of two site-pairs system where rest of the sites are also treated as environment.  }
\end{figure}
In Figure\,\ref{fig:figure1}(a), the schematic structure of FMO complex is shown where each sphere represents a BChl site.  Initial excitation occurs at site 1 or 6 and it is transported to sink sites 3 and 4 using sites 2, 5, and 7 as pathways in different possible configurations.

\subsection{Framework for simulating 2-sites system}
At the time of working with a specific site pair, we can treat the specific site pair with other sites as the system and the protein environment as the bath, or the specific site pair as the system and the protein environment with other sites as the bath (effective environment). In Figure \,\ref{fig:figure1}(b) we have shown two configuration of site-pairs where rest of the sites are also treated as environment along with protein environment to study the dynamics between the pairs. In the first case, the system will be represented by 3 qubits, and in the second case, the system will be represented by 1 qubit [as for representing $2^n -1$ sites, we need a minimum of n qubits]. If we treat the bath with minimum of 1 qubit, then in the first case we have to deal with 4 qubits or $2^4$ dimensional space, and in the second case we have to deal with 2 qubits or $2^2$ dimensional space. The second option has been chosen here to allow for less computational complexity.

For simulating site-pairs of FMO complex in discrete-time framework, we need to first initialize the state, follow it up with state evolution by taking all environmental effect into consideration and final state is measured. Before initializing the state,  we will demonstrate this problem using an illustration.
\begin{figure}[!h]
\centering
\includegraphics[width=0.3\textwidth]{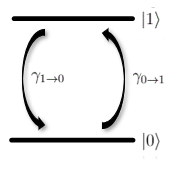}
\caption{\label{fig:figure2}A schematic illustration of two level system which is used to represent two sites where bath induces jump between the sites with probability  $\gamma_{i \rightarrow j}$.}
\end{figure}
In Figure\,\ref{fig:figure2}, the two level system is shown where the two levels are denoted by $\ket{0}$ and $\ket{1}$. The two sites of the FMO complex are mapped into these two levels. The bath induces a jump between these two sites. Here, the jump probability is denoted by $\gamma_{i \rightarrow j}$ where $i$ and $j$ are two sites. The initial system-environment state can be represented by
\begin{equation}\label{eq:1}
    \ket{\psi(0)}=\ket{i}_S\otimes \ket{0}_E.
\end{equation}
At the time of evolution, we need to capture the dissipative environmental effects as well as its internal unitary dynamics.
The dissipative effects can be represented in form of  environment-induced quantum jump which is implemented as
\begin{equation}\label{eq:2}
    \ket{i}_S\ket{0}_{E} \mapsto \sqrt{1-\gamma_{i\rightarrow j}}\ket{i}_S\ket{0}_{E} +\sqrt{\gamma_{i\rightarrow j}}\ket{j}_S\ket{1}_{E}.
\end{equation}
The evolution can be implemented as quantum channel by writing the Kraus operator representation in ref.\,\cite{base1} in matrix from. 
 The matrix form of the evolution operator can be written as
\begin{equation}\label{eq:3}
  \centering
M=M_1 C,
 \end{equation}
where $M_1$ and $C$ are both matrices given by, 
\begin{equation}\label{eq:4}
  \centering
M_1 = \begin{bmatrix}
\sqrt{1-\gamma_{i\rightarrow j}}    & 0 & 0 & 0  \\
   \sqrt{\gamma_{i\rightarrow j}}    & 0 & 0 & 0   \\
    0    & 0 & 1 & 0   \\ 
         0 & 0 & 0 & 1  
    \end{bmatrix},
\end{equation}
\begin{equation}\label{eq:5}
  \centering
C = \begin{bmatrix}
1   & 0 & 0 & 0  \\
   0    & 0 & 0 & 1   \\
    0    & 0 & 1 & 0   \\ 
         0 & 1 & 0 & 0  
    \end{bmatrix}.
\end{equation}
Here, our main goal is to focus on the dynamics between the two sites of FMO, so the coherent dynamics between two sites is implemented by,
\begin{equation}\label{eq:6}
    U (t) \equiv U_s(t) \otimes \mathbb{I},
\end{equation}
where
 \begin{equation}\label{eq:7}
   U_s(t)= e^{-i\frac{H_C}{\hbar}t},
 \end{equation}
 and
\begin{equation}\label{eq:8}
  \centering  
H_C= \begin{bmatrix}
    H_{11}       & H_{12}  \\
    H_{21}       & H_{22}
    \end{bmatrix}.
\end{equation}
The $\gamma$ values of Equation\,\ref{eq:2} and the Hamiltonian $H_C$ of Equation\,\ref{eq:8} have been taken from ref. \,\cite{data}, \cite{analysis1} and \cite{analysis2}.
The combined system and environment state is evolved as per Equation\,\ref{eq:3} and Equation\,\ref{eq:6} and the density matrix  at time $t$ will be,
\begin{equation}
\rho(t) = M U_s(t) \rho(0) U_s(t)^{\dagger} M^{\dagger}.
\end{equation}
By trace out the environment at each step we can see whether the reduced dynamics of system has any non-Markovian effects. 

\subsection{Trace distance and BLP measure}
The trace norm of a trace class operator $A$ is defined by
$||A|| = tr|A|$, where the modulus of the operator is given
by $|A| =\sqrt{A^\dagger A}.$ If $A$ is self-adjoint, the trace norm can be expressed as the sum of the moduli of the eigenvalues $a_i$
of A counting multiplicities, $$||A|| =\sum_i |a_i|.$$ This norm
leads to a natural measure for the distance between two
quantum states $\rho_1$ and $\rho_2$ known as trace distance,
\begin{equation}\label{eq:9}
D(\rho_1(t), \rho_2(t)) = \frac{1}{2}||\rho_1(t) - \rho_2(t)||.
\end{equation}
For pair of sites $i$ and $j$, $\rho_i(0) = |i\rangle \langle i |$  and $\rho_j (0)= |j\rangle \langle j |$. Trace distance is calculated for these two states as function of time.  Due to information flow between the open system and the environment we will see the change in trace distance with time.  If the information is lost to the environment from the system, we will only see a monotonic decrease in trace distance. However, along with the decrease if we observe some non-monotonic behaviour (small oscillations), it is considered as an indication of information back flow from environment to the system and  presence of non-Markovianity in the dynamics.  Small fluctuations in trace distance does not help us to quantify and compare the non-Markovian behaviour  between different site-pairs effectively.  Therefore, BLP measure which calculates the rate of change of trace distance between quantum states is used to quantify the non-Markovianity in the dynamics. The BLP measure is defined as follows,
\begin{equation}\label{eq:10}
\mathcal{N} = \int_{\sigma >0}\sigma(t,\rho_{i,j}(0))dt,
\end{equation}
where 
\begin{equation}\label{eq:11}
\sigma(t,\rho_{i,j}(0))=\frac{d}{dt}D(\rho_i(t),\rho_j(t)).
\end{equation}
According to this definition, the non-Markovian process will have a positive finite value for $\mathcal{N}$ where the value of the Markovian process of $\mathcal{N}$ will always be $0$ \cite{th1, review_nm, d_q_w}.  We will use trace distance to detect the presence of non-Markovianity in the dynamics. The BLP measure will help us to quantify the non-Markovianity between different site-pairs and identify the site-pairs with dominating non-Markovian behaviour.

\subsection{Framework for tunable bath couplings}

FMO dynamics depends on environmental assistance and in this framework, the environmental effects are implemented using quantum jumps. For the site-pair evolution,  Equation\,\ref{eq:2} is an implementation of a quantum jump, which happens due to the effect of environment on the system, and Equation\,\ref{eq:6} is an implementation of coherent dynamics which is due to couplings between different sites. For variable system-bath coupling, quantum jump rates will be affected by the coupling constant. Now in this scenario, the environment-induced quantum jump is implemented as follows:
\begin{equation}\label{eq:12}
    \ket{i}_S\ket{0}_{E} \mapsto \chi \sqrt{1-\gamma_{i\rightarrow j}}\ket{i}_S\ket{0}_{E} + \chi \sqrt{\gamma_{i\rightarrow j}}\ket{j}_S\ket{1}_{E},
\end{equation}
where $\chi \in [0,1]$. For this scenario, the modified matrix $M_1$ ($M'_1$) will take the form, 
\begin{equation}\label{eq:13}
  \centering
M'_1 = \begin{bmatrix}
\chi \sqrt{1-\gamma_{i\rightarrow j}}    & 0 & 0 & 0  \\
  \chi \sqrt{\gamma_{i\rightarrow j}}    & 0 & 0 & 0   \\
    0    & 0 & 1 & 0   \\ 
         0 & 0 & 0 & 1  
    \end{bmatrix}
\end{equation}
and the density matrix  at time $t$ will be,
\begin{equation}
\rho(t) = M'_{1} U_{s}(t) \rho(0) U_s(t)^{\dagger} M'^{\dagger}_{1}.
\end{equation}

\section{Results and Discussions}
\label{result}


Each site pairs of FMO complex have different features which facilitate the exciton energy transfer through specific paths. Thus, our primary goal is to check whether non-Markovianity or memory effects have any influence on the exciton energy transfer dynamics of the FMO complex. For doing that some selected site pairs are simulated using the framework presented in Sec.\,\ref{model_methods}. In addition to that, trace distance and BLP measure are calculated to detect the presence of non-Markovianity and quantify non-Markovianity, respectively in the dynamics. 

At the time of exciton energy transfer, it is transferred from site 1 or 6 to site 3, mainly through two dominant transfer pathways, which are $1 \rightarrow 2 \rightarrow 3$ and $6 \rightarrow (5,7) \rightarrow 4 \rightarrow 3$ as shown in Figure\,\ref{fig:figure1}(a). Thus, we have chosen 3 site pairs that are in the dominant transfer pathways (directly connected pairs) and other 3 site pairs that are not in the dominant transfer pathways to find the difference in the dynamics among the pairs.

\begin{figure}[!h]
\includegraphics[width=0.43\textwidth]{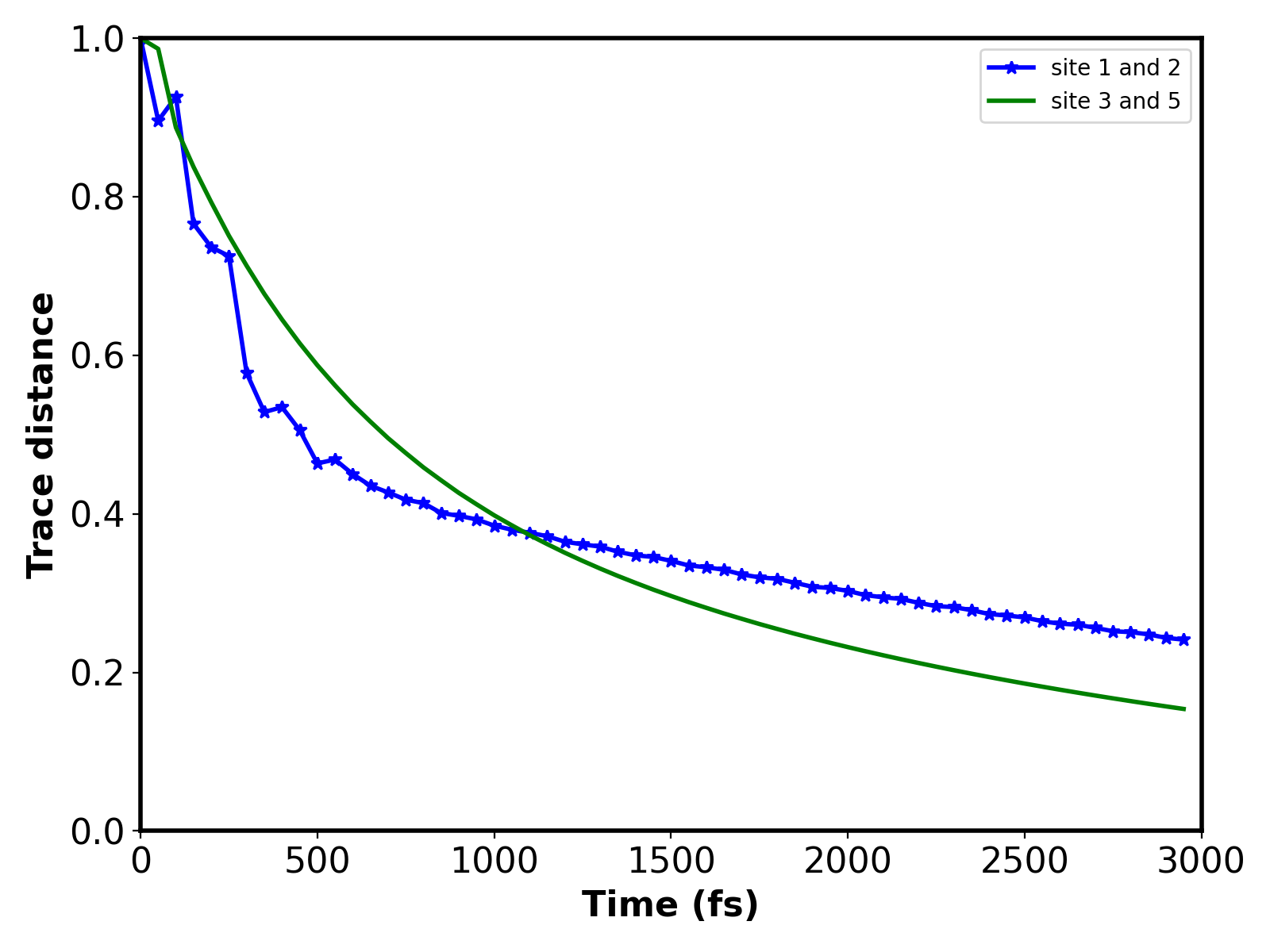}
\centering
 \caption{{\label{fig:figure3}Trace distance as function of time for site-pairs 1 - 2 and 3 - 5. For the directly connected sites (1-2) we can see small oscillation in trace distance whereas a monotonic decrease is seen for pair which has no direct connection.}}
\end{figure}
\begin{figure}[!h]
\includegraphics[width=0.43\textwidth]{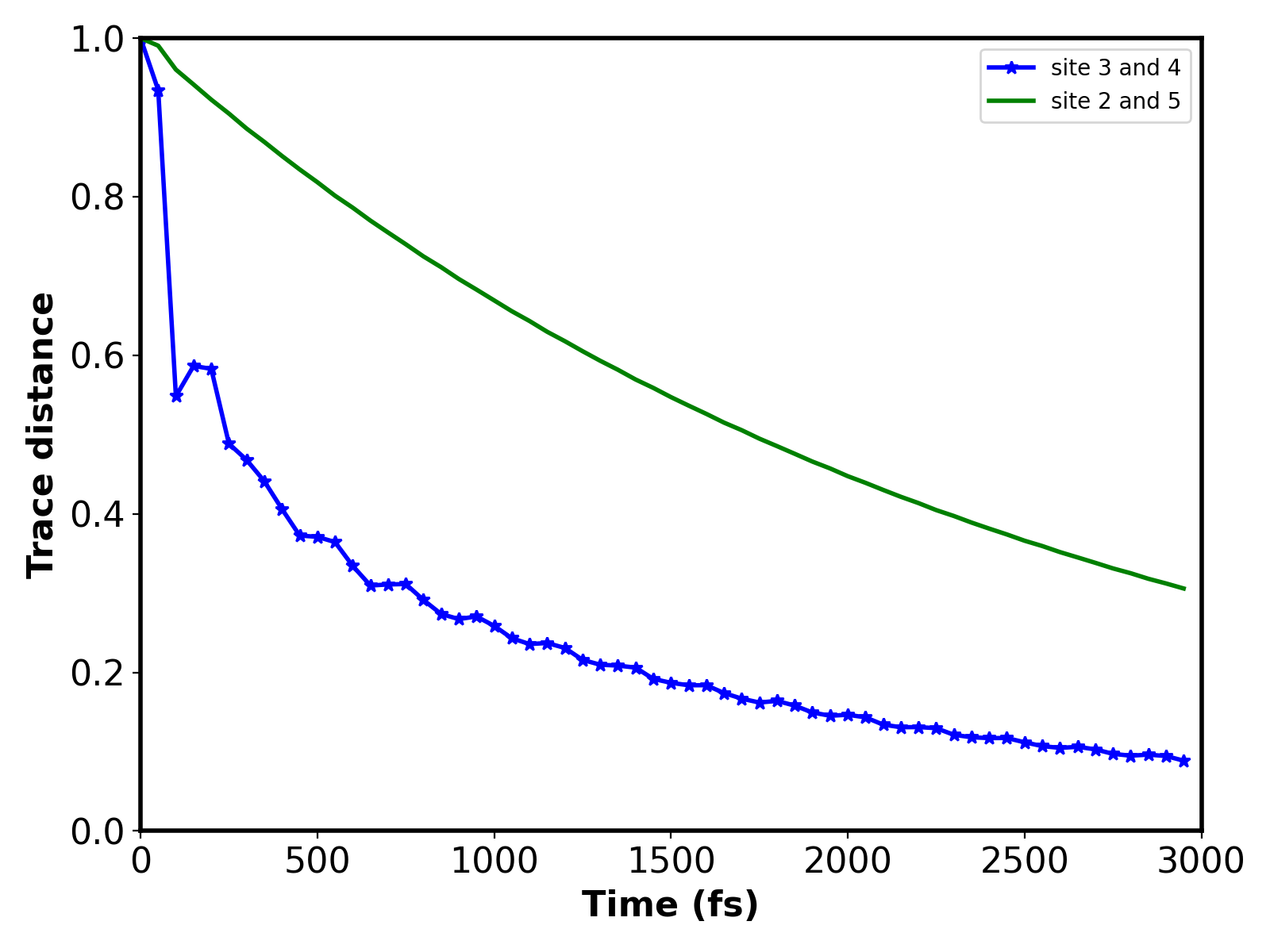}
\centering
\caption{{\label{fig:figure4}Trace distance as function of time for site-pairs 3 - 4 and 2 - 5. For the directly connected sites (3-4) we can see small oscillation in trace distance whereas a monotonic decrease is seen for pair which has no direct connection.}}
\end{figure}
\begin{figure}[!h]
\includegraphics[width=0.43\textwidth]{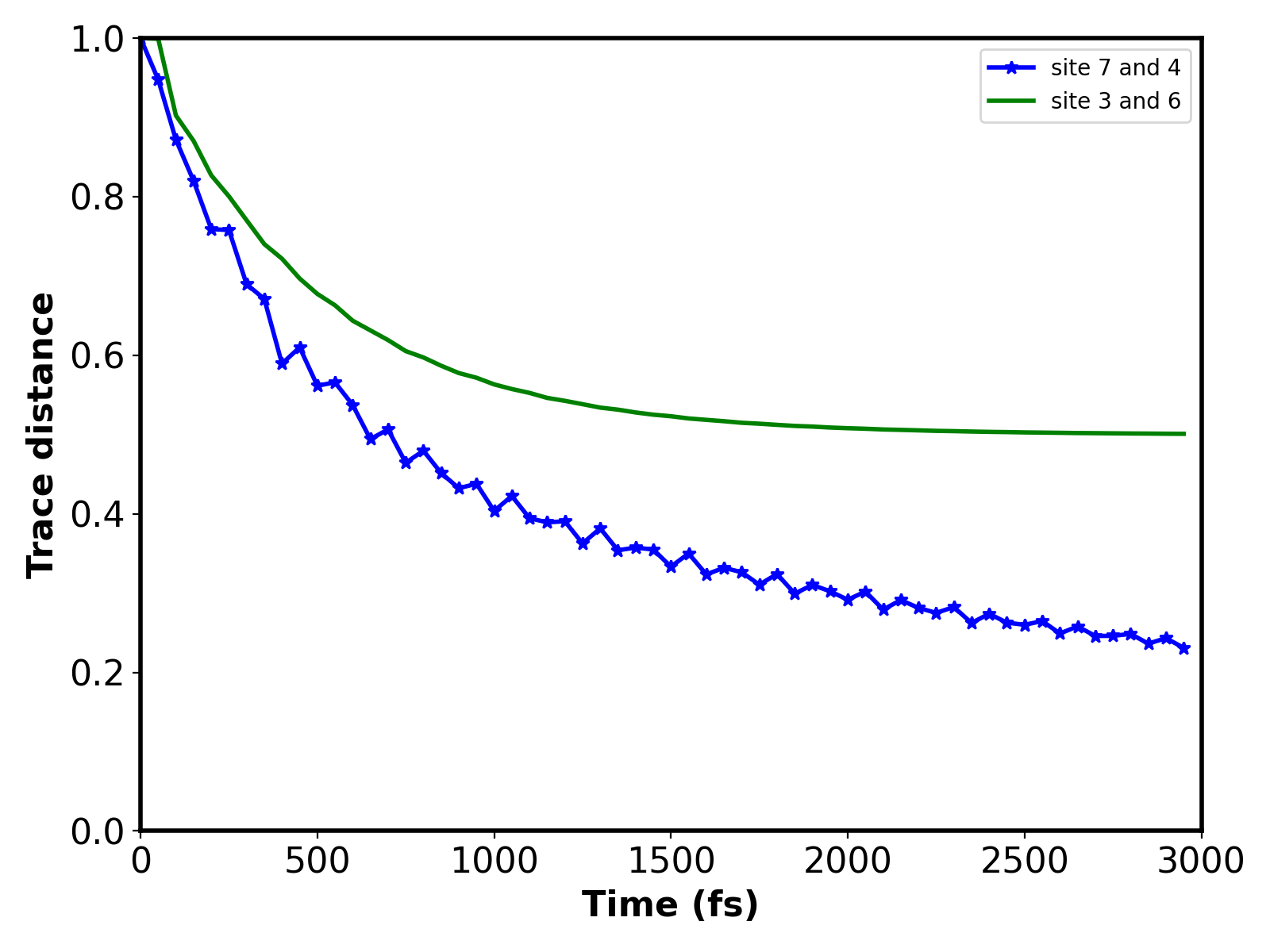}
\centering
 \caption{{\label{fig:figure5}Trace distance as function of time for site-pairs 4 - 7 and 3 - 6. For the directly connected sites (4-7) we can see small oscillation in trace distance whereas a monotonic decrease is seen for pair which has no direct connection.}}
\end{figure}
Figure\,\ref{fig:figure3}, Figure\,\ref{fig:figure4} and Figure\,\ref{fig:figure5}  presents the numerically calculated trace distance as function of time (fs) for site pairs with dominant transfer pathways, $(1, 2)$, $(3, 4)$ and $(7,4)$, and  site pairs  which are not in dominant transfer pathways $(3, 5)$, $(2, 5)$ and $(3,6)$.  The blue line curves with triangular symbol are for those site pairs present in the dominant transfer pathways and the green line curves are for those site pairs  which are not present in dominant transfer pathways. We see that for site pairs from dominant transfer paths the trace distance does not decrease monotonically like we see for site pairs which are not from dominant paths. For dominant transfer paths we can see small oscillations in trace distance with time, that is, $D(\Phi(\rho_1),\Phi(\rho_2))$ is increasing for some $t \ge 0$, which means that information is coming back into the system from environment. This indicates that those pair of sites have non-Markovianity in the dynamics.  The presence of non-Markovianity or memory effects in the dominant transport pathways of the FMO complex is an interesting result that is obtained here by simply considering site pairs in discrete-simulation framework and it clearly concur with previous theoretical understandings from earlier studies using hierarchical equations of motion (HEOM) approach\,\cite{non_mar}. Our site pair approach is computationally simplified discrete evolution that can be simulated using quantum circuits\,\cite{base1}.

\begin{figure}[!h]
\includegraphics[width=0.43\textwidth]{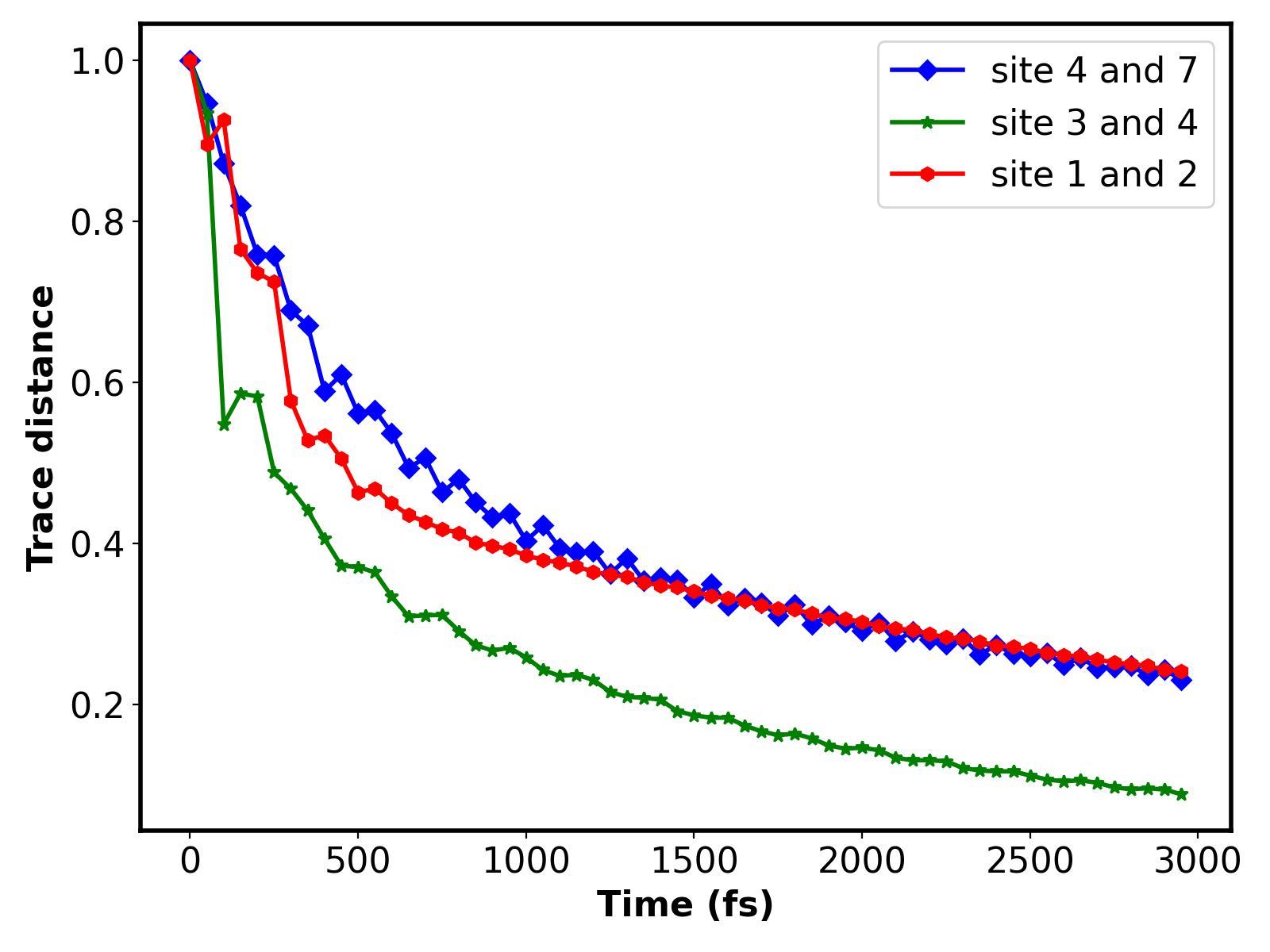}
\centering
 \caption{{\label{fig:figure6} Trace distance as function of time for the three dominant transfer pathways which show non-Markovianity. From the plots we can note that the trance distance measure does not allow us to clearly identify the dominating pathway. } }
\end{figure}

It can be understood that the site pairs present in the dominant transfer pathways have site coupling much higher than the others. Site coupling plays a very important role in exciton energy transfer because on the basis of site couplings the exciton de-localizes in between sites, which introduces a wavelike behaviour resulting in speedup of transport. From this result, we can say that the non-Markovianity which is present only in some specific site pairs depends on the site coupling, an internal feature of the system.  But from the trace distance measure we cannot quantify the non-Markovianity between the site pairs and explicitly identify the most prominent pairs along the dominant pathways.  In Figure\,\ref{fig:figure6}  we have shown the trance distance for the three site pairs with dominant transfer pathways and it is difficult to identify the most prominent pathway. Therefore,  BLP measure is used to explicitly quantified the presence of non-Markovianity for those site pairs with dominant couplings. For weakly coupled pairs, the trace distance decreases monotonically with time, thus the BLP measure is zero for such pairs according to Equation\,\ref{eq:10}.

\begin{figure}[!h]
\includegraphics[width=0.43\textwidth]{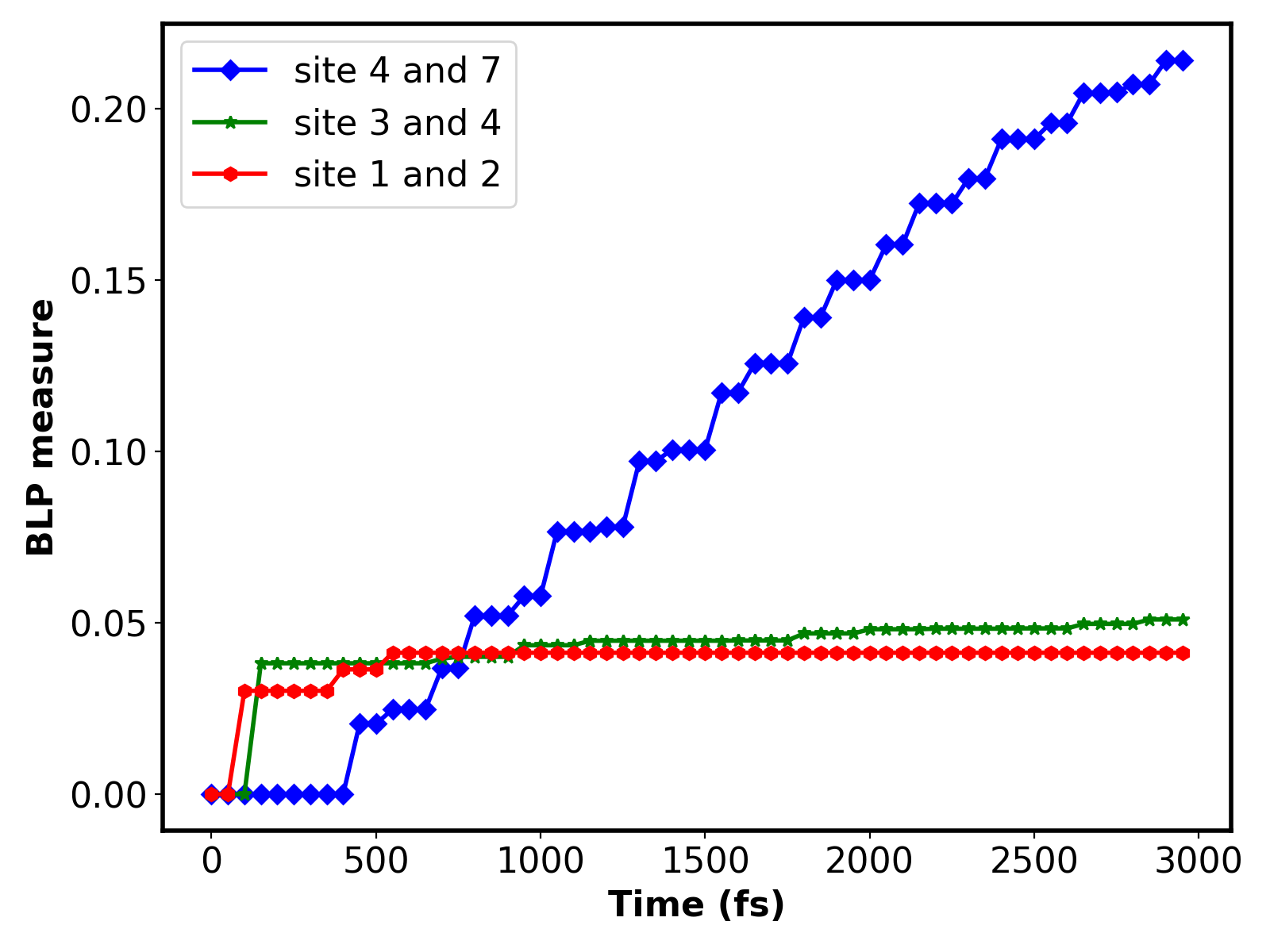}
\centering
 \caption{{\label{fig:figure7}BLP measure as function of time for different site pairs with dominant pathways.  The BLP measure quantifies the non-Markovianity and we can see that the non-Markovian oscialltion stay for long for site pair 4-7 compared to other pairs.}}
\end{figure}
Figure\,\ref{fig:figure7} shows the change of the BLP measure with time for all those sites where the trace distances do not decrease monotonically.
In the case of site-pairs 4-7, the non-Markovian oscillations stay longer than for site-pairs 1-2 and 3-4. As a result, in Figure\,\ref{fig:figure7} we can see that the BLP measure grows with time and also has a higher value. In the other two cases, the BLP measure has a lower value and almost saturates after a certain time. 

This result is consistent with other theoretical findings that the non-Markovianity decreases with increasing site energy difference. Site pair 4-7 has much lesser site energy difference than site pairs 1-2 and 4-7, as a result site pair 4-7 has higher non-Markovianity. If we look into the FMO complex, physical understanding is that, it possible for decreasing non-Markovianity with increasing site energy difference: as the energy difference increases, the site coupling becomes less significant and the sites become more independent resulting in decrease in non-Markovianity. Until now, we have explored non-Markovianity based only on the internal feature of the FMO complex, but the external environment plays an important role in the exciton energy transport dynamics, thus it is important to know how environmental effects make impact on non-Markovian behaviour.  The variable parameters in the discrete framework used in this works makes it easier to see such effects.

 \begin{figure}[!h]
\includegraphics[width=0.43\textwidth]{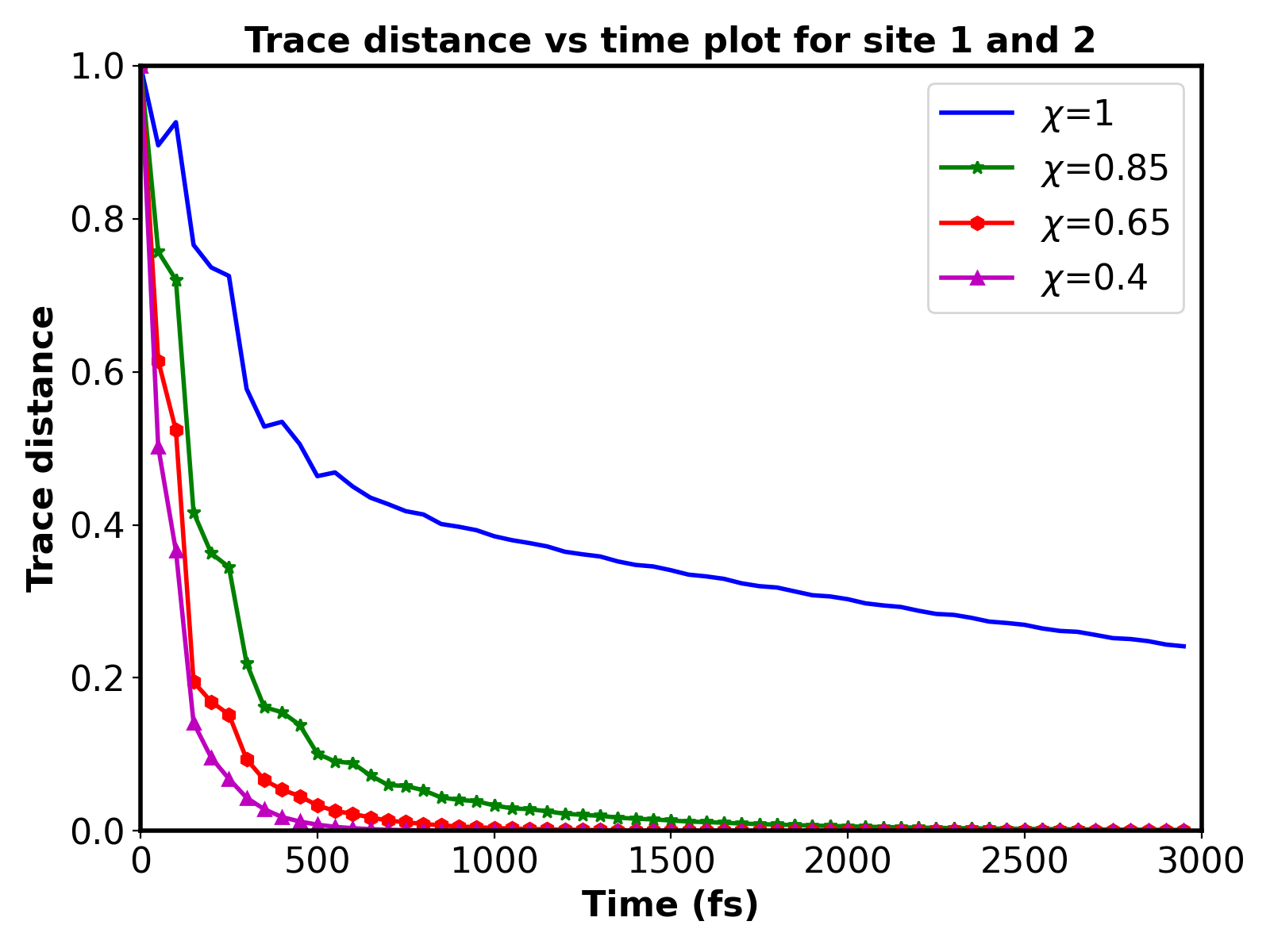}
\centering
 \caption{{\label{fig:figure8}Trace distance as function of time for site pair 1-2 with tunable system-bath couplings. With increase in coupling strength we see an increase in non-Markovian oscillation.}}
\end{figure}
\begin{figure}[!h]
\includegraphics[width=0.43\textwidth]{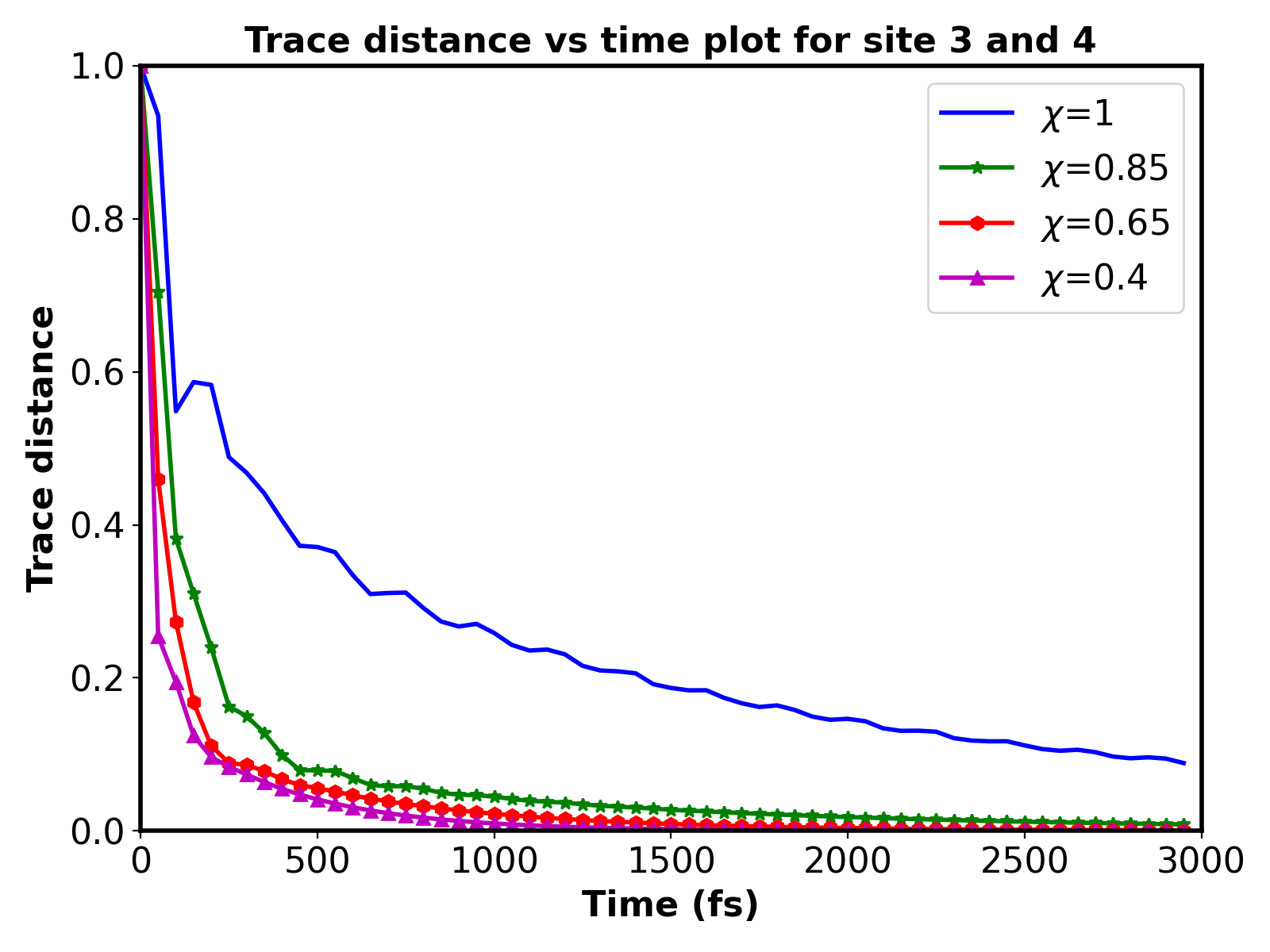}
\centering
 \caption{{\label{fig:figure9}Trace distance as function of time for site pair 4-3 with tunable system-bath couplings. With increase in coupling strength we see an increase in non-Markovian oscillation.}}
\end{figure}
\begin{figure}[!h]
\includegraphics[width=0.43\textwidth]{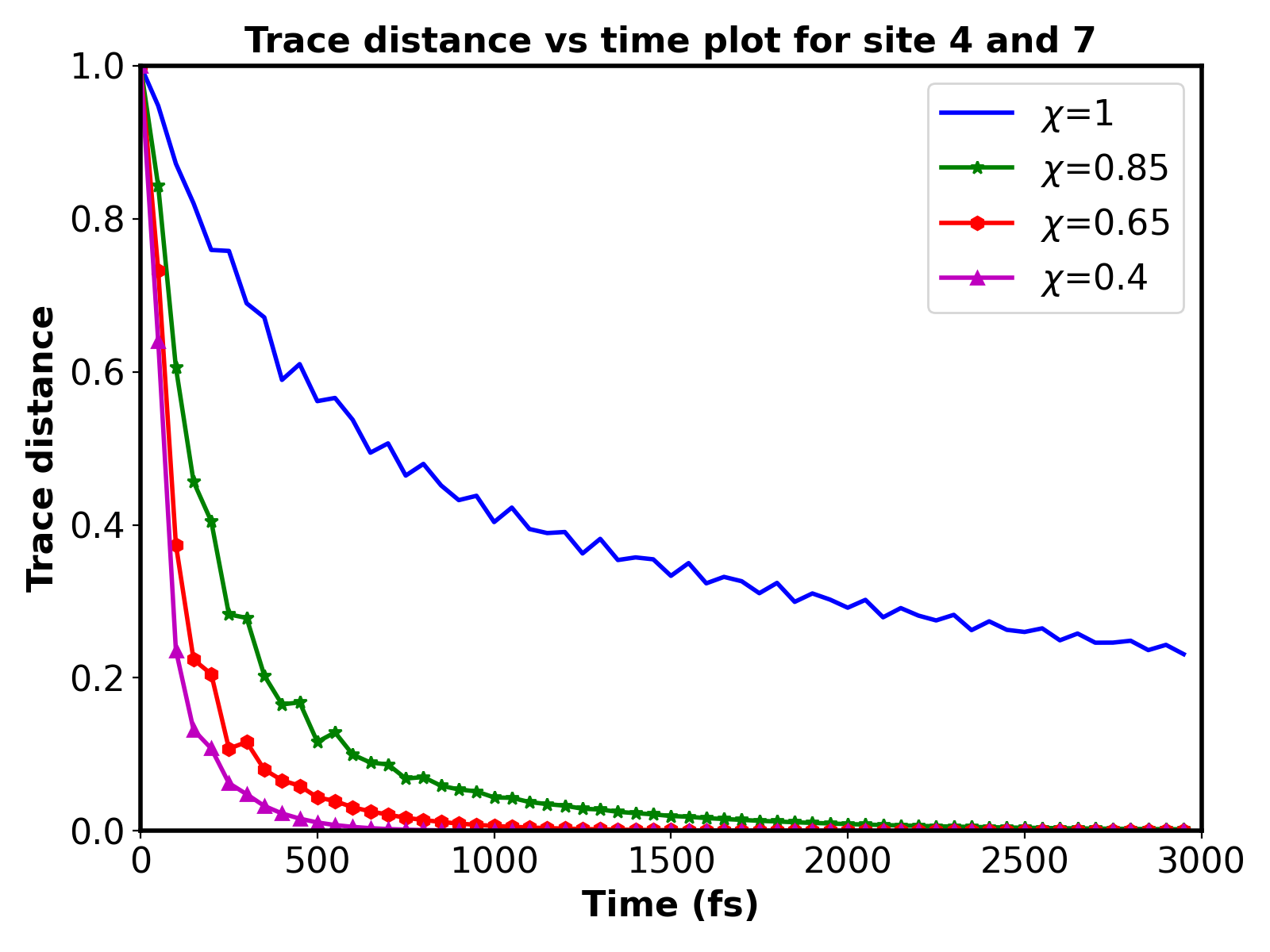}
\centering
 \caption{{\label{fig:figure10}Trace distance as function of time for site pair 4-7 with tunable system-bath couplings. With increase in coupling strength we see an increase in non-Markovian oscillation.}}
\end{figure}

Figure\,\ref{fig:figure8}, Figure\,\ref{fig:figure9} and Figure\,\ref{fig:figure10} show the trace distance as function of time with different system-bath coupling strength for different site-pairs, 1-2, 3-4, and 4-7, respectively. The values are calculated for four different coupling constants ranging from 1 to 0.4. One common conclusion can be drawn from all these plots is that the weakly coupled environment leads to less non-Markovian oscillation than the strongly coupled one. In case of strong system-bath couplings, environment is having more influence on the system, thus the lost information from system is getting stored in the environment for some time and again coming back  into the system after some time. Thus, we observe more non-Markovian oscillations in case of strong system-bath couplings.

\begin{figure}[!h]
\includegraphics[width=0.42\textwidth]{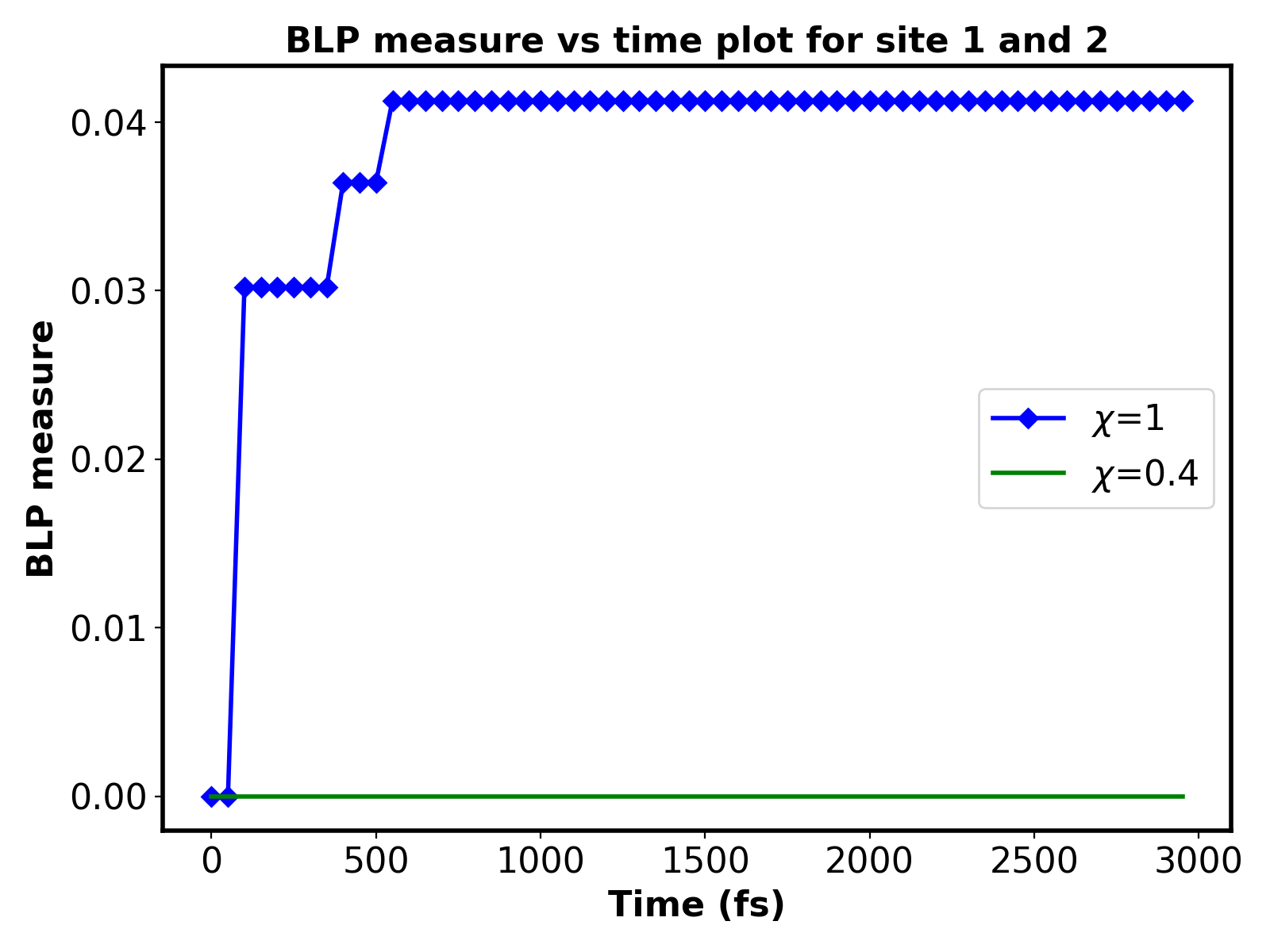}
\centering
 \caption{{\label{fig:figure11}BLP measure as function of time with strong and weak system-bath couplings for site pair 1-2. No information back-flow is seen for weak coupling.}}
\end{figure}
\begin{figure}[!h]
\includegraphics[width=0.42\textwidth]{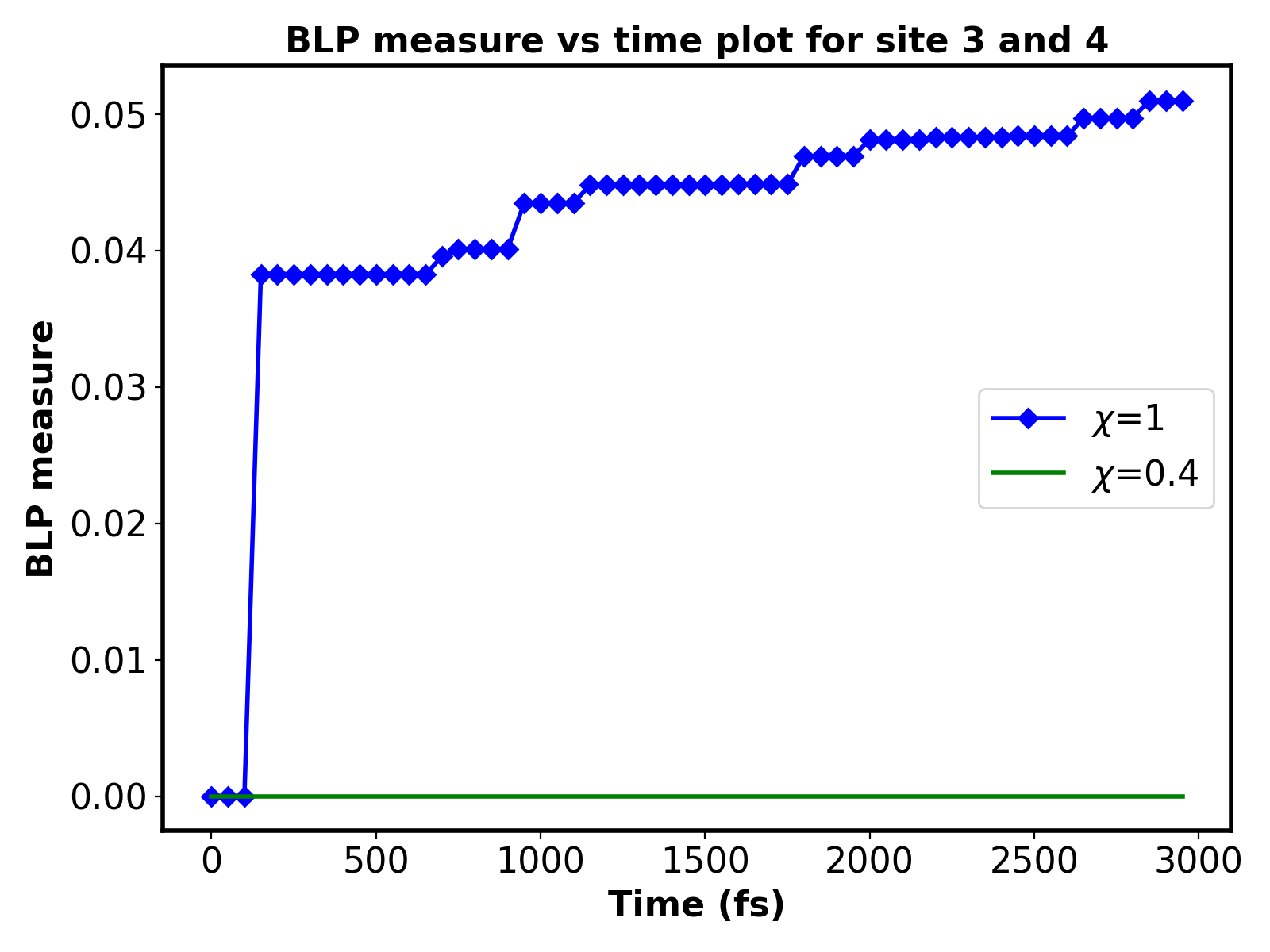}
\centering
 \caption{{\label{fig:figure12}BLP measure as function of time with strong and weak system-bath couplings for site pair 3-4. No information back-flow is seen for weak coupling.}}
\end{figure}
\begin{figure}[!h]
\includegraphics[width=0.42\textwidth]{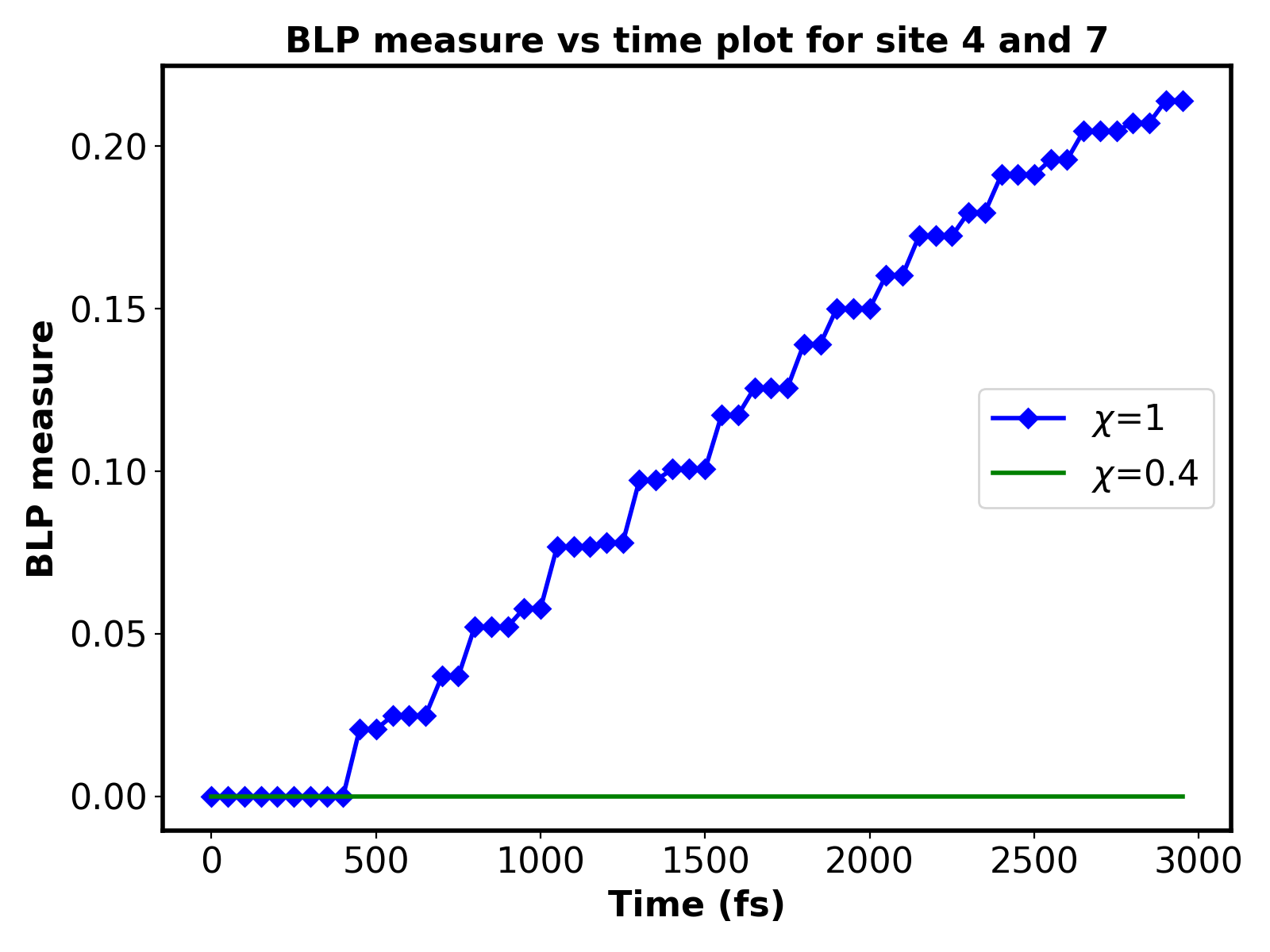}
\centering
 \caption{{\label{fig:figure13}BLP measure as function of time with strong and weak system-bath couplings for site pair 4-7. No information back-flow is seen for weak coupling.}}
\end{figure}

Next, we have quantified non-Markovianity using BLP measure for these site pairs in the case of coupling constant 1 (high level coupling) and 0.4 (low level coupling).

Figure\,\ref{fig:figure11}, Figure\,\ref{fig:figure12} and Figure\,\ref{fig:figure13} shows the BLP measure as function of time plot for the site pairs  1-2, 3-4, and 4-7, respectively. From the observations we can conclude that in case of low level system-bath coupling (coupling constant=0.4), BLP measure is zero which denotes no information back-flow into system indicating that it is a Markovian dynamics, where in case of high level system-bath coupling (coupling constant=1.0) BLP measure gives non-zero value which denotes information back-flow into the system from environment indicating the existence of non-Markovianity in dynamics. Thus, here we can see that by using variable system-bath couplings, it is possible to tune and interchange between non-Markovian and Markovian dynamics. It is also important to note here that the system-bath couplings depend on temperature, thus by tuning temperature it is possible to interchange between Markovian and non-Markovian dynamics.  In our study the dynamics between site-pairs of the FMO complex were numerically simulated using discrete-time quantum  jumps model on classical computer. The information about their dynamics and the presence of non-Markovianity has been calculated  using trace distance and the BLP measure method. From our results, it can be established that the presence of non-Markovianity in some specific site pairs depends on the internal structure of the FMO complex and the influence of the environment.  The model also shows that the specific site-pair dynamics and environment can be simulated using two qubit quantum simulator.

\section{Conclusions}
\label{conc}

We have presented a theoretical framework for using environment driven quantum jump model to study  site-pair dynamics of FMO complex.  The site-pair description along with tunability of system-environment coupling presented in this work has paved way to engineer the dynamics and explore the presence of quantum features across variable coupling parameters.  
We have used trace distance to establish the presence of non-Markovian memory effects and  BLP measure to quantify the memory effect between site pairs with variable  system-environment coupling strength.  This memory effect is controlled by structural features of the FMO complex (site couplings and site energy differences) and environmental influence. Our results using discrete-time approach match with previous theoretical findings using continuous-time approach and show that non-Markovian memory effects is present in those conditions where internal structures and environmental effects are in favor of faster transport. Unlike in previous results,  using BLP measure we could also quantify the BLP measure and establish that the higher degree of non-Markovianity results in dominating pathways for faster transport. BLP measure could also show that the memory effect is zero even for dominant pathways when the system-environment coupling is weak.  Therefore, critical coupling strength is important to see memory effect even for the prominent pathways.

We may conclude from these results that the non-Markovian memory effects facilitate energy transfer by giving special importance to certain site pairs that help the exciton to travel through a certain direction from antenna to sink. This study also sheds light on the fact that memory effects can help in noise-assisted transport dynamics, which can be useful for real-life applications.  The simple site pair unitary dynamics and environment driven quantum jump model presented here is ideal for discrete (digital) quantum simulation and provides universality. Therefore,  the same kind of framework can be applied to model and study the dynamics in other chemical complexes. \\

\noindent
{\bf AUTHOR INFORMATION :}\\

\noindent
{\bf Corresponding Author} \\
{\bf C. M. Chandrashekar -}  Quantum Optics \& Quantum Information,  Department of Instrumentation and Applied Physics, Indian Institute of Science, Bengaluru 560012, India ;  The Institute of Mathematical Sciences, C. I. T. Campus, Taramani, Chennai 600113, India ;  Homi Bhabha National Institute, Training School Complex, Anushakti Nagar, Mumbai 400094, India ;  https://orcid.org/0000-0003-4820-2317 ;   Email  : chandracm@iisc.ac.in \\

\noindent
{\bf Author }\\
{\bf Mousumi Kundu - }Indian Institute of Science Education and Research, Berhampur\\

\noindent
{\bf  Notes}\\
The authors declare no competing financial interest.\\
 
 \noindent
 {\bf Acknowledgment :} We acknowledge support from the Interdisciplinary Cyber Physical Systems (ICPS) Programme of the Department of Science and Technology, Government of India. Grant No. DST/ICPS/QuST/Theme-1/2019.


\end{document}